\documentclass{article}

\usepackage{enumerate}
\usepackage{fullpage}
\usepackage{amsfonts} 
\usepackage{mathtools}
\usepackage{units} 
\usepackage{amsthm}	
\usepackage{placeins}
\usepackage{float}
\usepackage{amsthm}
\usepackage{framed}
\usepackage[english]{babel}

\newcommand{\appr}[1]{\textbf{Approximation \ref{#1}}}

\newtheorem{approximation}{Approximation}

\begin{document}
\title{Fringe Field Effects on Bending Magnets, Derived for TRANSPORT/TURTLE}
\author{Riley Molloy and Sam Blitz}
\maketitle

\begin{abstract}
A realistic magnetic dipole has complex effects on a charged particle near the entrance and exit of the magnet, even with a constant and uniform magnetic field deep within the interior of the magnet. To satisfy Maxwell's equations, the field lines near either end of a realistic magnet are significantly more complicated, yielding non-trivial forces. The effects of this fringe field are calculated to first order, applying both the paraxial and thin lens approximations. We find that, in addition to zeroth order effects, the position of a particle directly impacts the forces in the horizontal and vertical directions.
\end{abstract}

\section{Introduction}

Charged particle optics is frequently used within the context of experimental beamline design. A key element within an optical system is often a magnetic dipole, used to bend a beam. A pure dipole field implies that the trajectory of charged particles will strictly follow an arc. However, a physical implementation of such a dipole necessitates effects beyond the scope of a pure dipole field. In the FNAL experiment ORKA \cite{proposal}, dipoles are used for typical reasons such as spatial constraints and the removal of certain unwanted particles. To reduce the length of the beamline (and therefore minimzing travel time), designer Jaap Dornboss suggested the implementation of a dipole adjusted at its front and back to introduce additional focusing and defocusing effects on the beam in both the vertical and horizontal directions. These effects are a direct result of the fringe fields, or the non-constant fields located at the ends of the dipole magnet. It is important to carefully consider these effects because of the sensitive nature of the ORKA secondary beamline to subtle differences in (specifically vertical) beam properties.

This document provides an in-depth derivation of the transfer matrix elements of the fringe field of a dipole magnet. In particular, standard optical approximations are applied to achieve a ``first order" approximation of the fringe field. Initally, the system is simplified to consider an infinitesimally thin fringe field called the ``sharp cutoff field." Then, we consider a finite-length fringe field and its effects compared to that of the sharp cutoff field. Throughout the document, we assume conventional terminology and notation used in beam transport and as established in Refs.~\cite{dave} and \cite{enge}. We also assume some basic knowledge of E\&M and mechanical knowledge of a dipole magnet's structure.

Throughout this document, we emphasize the approximations used to arrive at the solutions. For many magnets, these approximations yield accurate results; however, it is up to a designer to determine whether the approximations used are sufficiently accurate for consideration of the fringe field. We find that magnetic dipoles tuned for low momentum particles ($<1 \, \mathrm{GeV}/c$), bent through large angles ($>1\,  \mathrm{rad}$), and/or containing large vertical gaps relative to their lengths require a higher-order consideration of fringe field effects for accurate analysis.

\section{Impulse Approximation of Magnet Exit}
In this section, we look to simplify the system by considering the effects of a fringe field with zero length. This configuration implies a constant field inside the magnet until the defined field boundary, after which the magnitude immediately drops to zero. 

\subsection{Coordinate Systems}

\begin{figure}[H]
\centering \includegraphics[scale=.6]{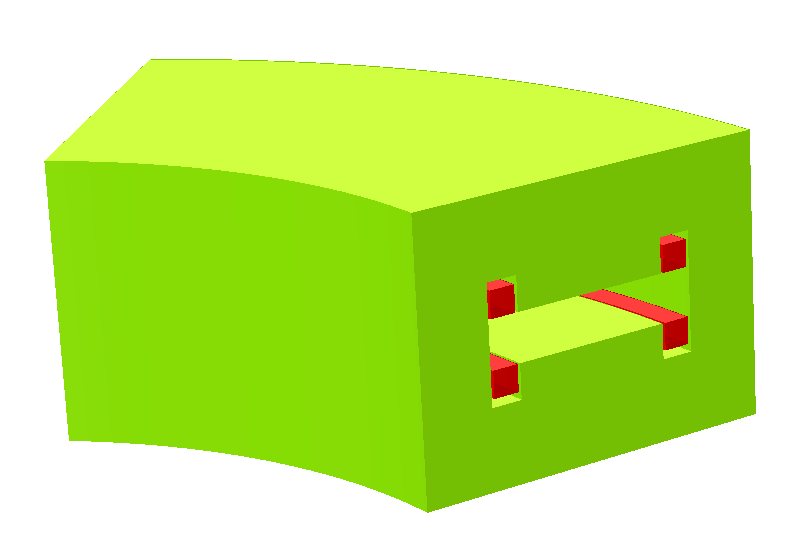} 
\caption{A 3-dimensional rendering of a simplified yet geometrically realistic dipole bending magnet. In this graphic, the ``pole faces" are the two rectangular faces between the coils. The region of interest is a thin mid-plane volume.}
\label{3d-pic}
\end{figure}

When discussing the fringe field, calculations are simplified by placing the origin of each coordinate system at the center of the pole face. The natural coordinate system defined throughout the magnet is a cylindrical curvilinear coordinate system based on the radius of curvature of the reference trajectory\footnote{The reference trajectory is the uniform trajectory of the central ray. All other rays are measured relative.} given by the  magnetic field within the magnet. This system is called the $(x,y,t)$ coordinate system.

\begin{figure}[H]
\centering \includegraphics[scale=.35]{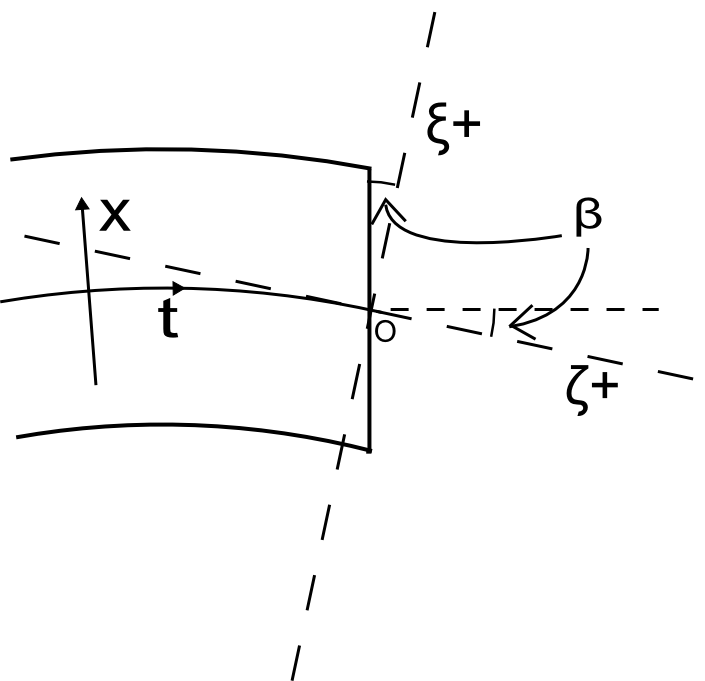} 
\caption{A description of the coordinate system. Note that the beam travels from left to right, with $t-$ and $\zeta-$ inside the magnet. A positive $\beta$ implies a counter-clockwise rotation of the pole face when viewed from above.}
\label{coordsys}
\end{figure}

A new coordinate system is introduced to describe the motion of a particle upon exiting the magnet.  This coordinate system again has an origin centered on the pole face and extends rectilinearly beyond the magnet, following the reference trajectory. This coordinate system is called the $(\xi, y, \zeta)$ coordinate system. Note that the $y$ axis, in both coordinate systems, points out of the page in Fig.~\ref{coordsys}. 

Fig.~\ref{coordsys} depicts the new coordinate definitions based on the natural curvilinear coordinate system. In the curvilinear coordinate system, $t+$ follows the path of the reference trajectory through the magnet, and the $x$-axis points in the direction perpendicular to the $t$ axis in the bend-plane. Sometimes, a magnet's pole face is rotated about the $y$ axis through an angle $\beta$, conventionally positive for counter-clockwise rotations when viewed from above, to modify the effects on the beam.

\begin{figure}[H]
\centering \includegraphics[scale=.7]{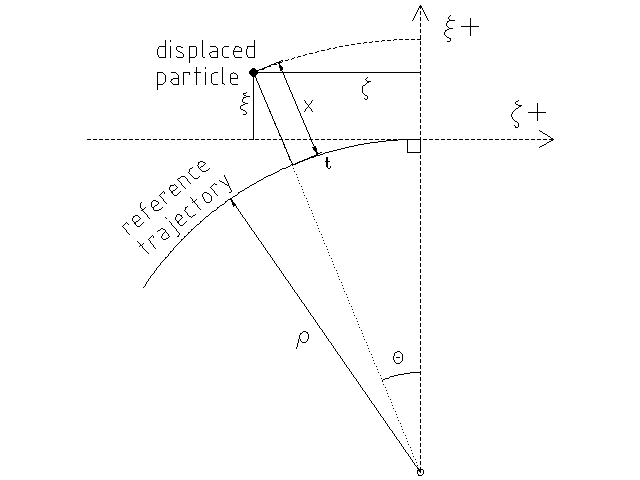} 
\caption{The trigonometry used to convert between the two coordinate systems. Note that both the angle $\theta$ and the quantity $t$ are both negative.}
\label{coordtransform-pic}
\end{figure}

We first look to find an expression of $x$ and $t$ in terms of $\xi$ and $\zeta$ from Fig.~\ref{coordtransform-pic}. Figure~\ref{coordtransform-pic}) shows the region surrounding the pole face and a displaced particle with its corresponding position in both $(x,y,t)$ and $(\xi, y, \zeta)$. This image is centered around the origin of both coordinate systems (located at the pole face). Note that this graphic does not show the geometry of the magnet. For this reason, it should only be used for conceptualizing the relationship between coordinate systems.

\begin{figure}[H]
\centering \includegraphics[scale=.25]{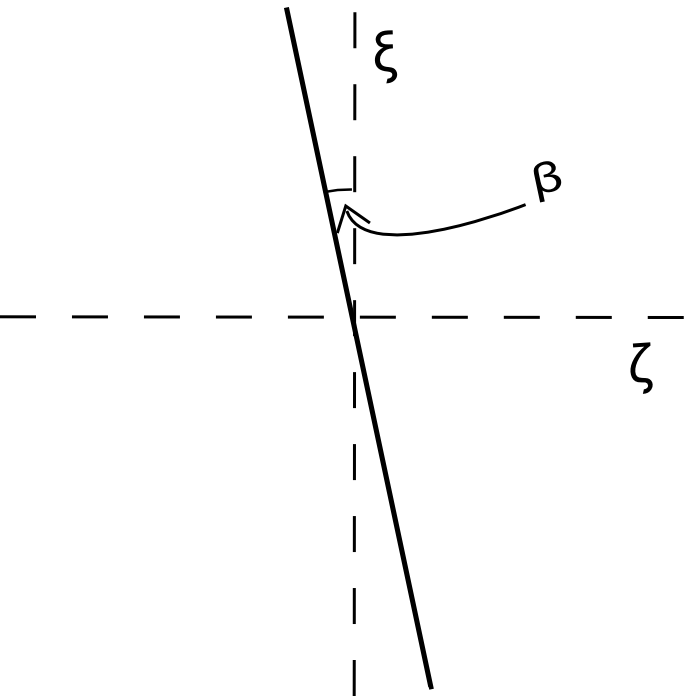}   
\caption{A plot of the pole face line in the $\xi \zeta$ plane.}
\label{poleface-equation}
\end{figure}

From the pythagorean theorem, we find that $(\rho + \xi)^2 + \zeta^2 = (\rho + x)^2,$
or \\ $x = \sqrt{\zeta^2 + (\rho + \xi)^2} - \rho$. To solve for $t$, we recognize that $t = \rho \theta$, where $\theta$ is the angle traversed in the path and $\rho$ is the radius of curvature. Then, $\tan \theta = \frac{\zeta}{\rho + \xi}$. Together,
\begin{align} \label{coordtransform}
& x = \sqrt{\zeta^2 + (\rho + \xi)^2} - \rho \notag \\
& t = \rho \, \mathrm{arctan} \frac{\zeta}{\rho + \xi}.
\end{align}
Solving for $\zeta$ and $\xi$, we have 
\begin{align}
& \xi = \rho \left( \cos \frac{t}{\rho} - 1 \right) + x \cos \frac{t}{\rho} \notag \\
& \zeta = (\rho + x) \sin \frac{t}{p} \label{xizeta}
\end{align}
Now we construct an expression for the line that defines the exit pole-face in the $\xi \zeta$ coordinate system as seen in Fig.~\ref{poleface-equation} with 
\begin{equation} \label{xitanbeta}
\zeta = - \xi \tan \beta,
\end{equation}
where $\beta$ is the pole-face rotation angle.

The component perpendicular to the bend plane, $y$, is identical in both coordinate systems. We retain the same coordinate name for clarity and convenience. A point in the curvilinear coordinate system is given by $(x,y,t)$ while a point in the rectilinear coordinate system is given by $(\xi, y, \zeta)$. 

\subsection{Horizontal Effects}

We begin the process of calculating the horizontal focusing effect of the fringe field. The initial position of the particle is given as $(x_0,y_0,0)$ in the $(x,y,t)$ coordinate system. This particle lies on the ``reference plane" which is both the $xy$ plane and the $\xi y$ plane at $t=0$ and $\zeta=0$ respectively. This plane coincides with the pole face when $\beta=0$. 

To calculate the horizontal effects, we must first trace the trajectory of a test particle\footnote{We designate an arbitrary particle off the reference trajectory as a ``test particle." It is distinct frommm the reference particle, which lies along the reference trajectory. This particle may or may not differ slightly in momentum from the reference particle.} backwards in the $(x,y,t)$ coordinate system from the reference plane until it intersects the rotated pole-face. We then change to the rectilinear $(\xi,y, \zeta)$ coordinate system so that we can translate the particle through the wedge-shaped drift space. The net effect of this transformation removes the field's effect on the particle through the region swept out by a rotation of the pole face. (Note that while this description assumes positive $\beta$ and $x_0$, the processes used can be rearranged for any condition. The final results are always the same.)

We know from established TRANSPORT documentation that the transformation through a dipole with radius of curvature $\rho$ and vertical magnetic field derivative $n$ (a unitless value equal to zero in the case of any relevant dipole magnet) is given as 
\begin{align}
x = x_0 c_x(t) + x'_0 s_x(t) + \delta d_x(t) \notag \\
x' = x_0 c'_x (t) + x'_0 s'_x(t) + \delta d'_x(t) \notag
\end{align}
where $c_x(t) = \cos (k_x t)$, $s_x(t) = \frac{1}{k_x} \sin (k_x t)$, \\ $d_x(t) = \frac{1}{k_x^2 \rho} (1- \cos(k_x t))$, and $k_x^2 = \frac{1-n}{\rho^2}$. Note that the prime indicates differentiation with respect to the longitudinal axis---in this case, $t$---for the remainder of this document. Expanding the trajectories about $t=0$, we have
\begin{align}\label{expanded x from magnet}
x (t) &= x_0 \left(1 - \frac{k_x^2 t^2}{2} + \dots\right) + \frac{x_0 '}{k_x} \left(k_x t - \frac{k_x^3 t^3}{6} + \dots\right) \notag \\ &+\frac{\delta}{ k_x^2 \rho} \left(1 - 1 + \frac{k_x^2 t^2}{2} - \dots\right)
\end{align}
\begin{align}\label{expanded x' from magnet}
x'(t) &= -x_0 k_x \left(k_x t -\frac{k_x^3 t^3}{6} + \dots\right) + x_0 ' \left(1 - \frac{k_x^2 t^2}{2} + \dots\right) \notag \\&+ \frac{\delta}{k_x \rho} \left(k_x t - \frac{k_x^3 t^3}{6} + \dots\right).
\end{align}
Note: We define $\delta$ as the percentage difference between the test particle's momentum and the reference momentum.

\begin{framed}
\begin{approximation}\label{paraxial approximation}
From the paraxial approximation used in first-order optics, we assert that each ray makes a small angle with the reference trajectory and lies a small distance from the reference ray compared to the radius of curvature of the reference trajectory. In the $(x,y,t)$ coordinate system used, this assumption implies
$$\frac{x}{\rho} \ll 1, \; \;  \frac{y}{\rho} \ll 1$$
$$x' \ll 1, \; \; y' \ll 1, \; \; \delta \ll 1$$
The value of $\delta$ is constrained to be significantly less than one in order to ensure that the angles formed with the reference trajectory stay small.
\vspace{.1in} 

\noindent Because these values are small, any product of two or more of these quantities can be neglected in the first-order approximation.
\end{approximation}
\end{framed}

\begin{figure}[H]
\centering \includegraphics[scale=1]{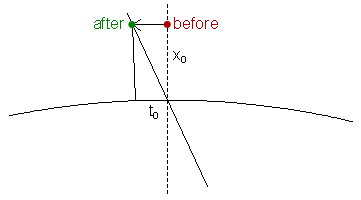} 
\caption{A graphical representation of a ray traced backwards from the reference plane. Note that $t_0$ decreases as $x_0$ decreases.}
\label{t0pic}
\end{figure}

In order to transform the particle backwards through the magnetic field (described above), we evaluate $x(t_0)$, $x'(t_0)$, where $t_0$ is the traversed longitudinal displacement from the reference plane to the pole face (see Fig.~\ref{t0pic}).  Applying geometery and $x_0 / \rho \ll 1$ from \appr{paraxial approximation}, we observe that $t_0/ \rho \ll 1$. We can therefore treat it as a small quantity as we do with $x_0 / \rho$. In fact, dividing both sides of Eq.~\eqref{expanded x from magnet} by $\rho$, we recognize there are several terms that contain a product of $t_0 / \rho$ and some quantity mentioned in \textbf{Approximation \ref{paraxial approximation}}. Eliminating those terms and again multiplying through by $\rho$, we have $x=x_0$. Applying the same argument to Eq.~\eqref{expanded x' from magnet}, we eliminate the products of small terms, yielding $x' = x_0'$. Together,
\begin{align}
x &= x_0 \notag \\
x' &= x_0'. \label{initxx'}
\end{align}

The above calculations show that tracing the particle backwards from the reference plane to the pole-face yields no first order changes. 

The rotated pole-face geometrically yields a focusing or defocusing effect on horizontally displaced particles. Physically, the focusing effect is simply the result of passing through less (or more) field region. If a particle has a positive $x$ coordinate and the pole-face rotation angle $\beta$ is positive, then it will experience less field than the reference trajectory; likewise, if a particle has a negative $x$ coordinate and $\beta$ is positive, it will experience \emph{more} field than the reference trajectory. The reverse occurs for a negative $\beta$. Thus, we can expect that a positive $\beta$ results in a defocusing effect (the horizontal exit slope of particles with negative $x_0$ decreases and the horizontal exit slope of particles with positive $x_0$ increases) and similarly a negative $\beta$ results in a focusing effect.

\begin{figure}[H]
\centering \includegraphics[scale=.4]{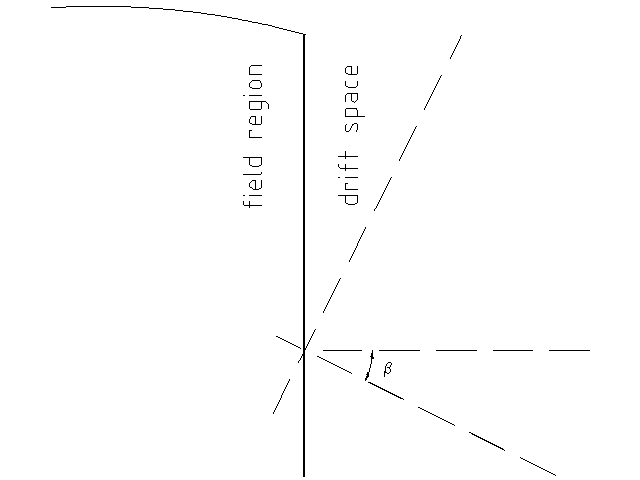} 
\caption{Graphical representation of the field boundary, the wedge-shaped drift space, and the refrence plane.}
\label{wedgeds}
\end{figure}

To express this behavior mathematically, we must change to the $\xi \zeta$ coordinate system to reflect the wedge-shaped, field-free region outside of the pole-face (see Fig.~\ref{wedgeds}). This change in coordinate system reflects the physical fact that the particle's direction does not change after exiting the field region: any ray will travel in a straight line once in the rectilinear coordinate system.

By Eq.~\eqref{xizeta}, we recognize that when we expand $\xi$ in $t$ and evaluate it at $t=t_0$, \textbf{Approximation \ref{paraxial approximation}} yields $\xi = x$. Because $x = x_0$ to first order at the pole face (see Eq.~\eqref{initxx'}), we conclude that, at the reference plane in the $\xi \zeta$ coordinate system, $\xi = x_0$. Therefore, we only look for effects in $\xi' = \frac{d \xi}{d \zeta}$ due to the change in coordinate system.

To find these effects, we begin by applying the chain rule:
\begin{align}
\frac{d \xi}{d \zeta} = \frac{\frac{d\xi}{dt}}{\frac{d\zeta}{dt}} = \frac{\frac{\partial \xi}{\partial x} \frac{dx}{dt} + \frac{\partial \xi}{\partial t}}{\frac{\partial \zeta}{\partial x} \frac{dx}{dt} + \frac{\partial \zeta}{\partial t}} \notag.
\end{align}
From Eq.~\eqref{xizeta}, we substitute in the appropriate derivatives to obtain
\begin{equation}
\frac{d \xi}{d \zeta} = \frac{ \frac{dx}{dt} \cos \frac{t}{\rho} - \left( 1 + \frac{ x}{\rho}\right) \sin \frac{t}{\rho} } {  \frac{dx}{dt} \sin \frac{t}{\rho} + \left( 1 + \frac{ x}{\rho}\right) \cos \frac{t}{\rho} } = \frac{ x_0' \cos \frac{t}{\rho} - \left( 1 + \frac{ x_0}{\rho}\right) \sin \frac{t}{\rho} } {  x_0' \sin \frac{t}{\rho} + \left( 1 + \frac{ x_0}{\rho}\right) \cos \frac{t}{\rho} } \notag.
\end{equation}
To simplify, we first expand in $t$ and evaluate at $t=t_0$, giving
\begin{equation}
\frac{d \xi}{d \zeta} = \frac{x_0' \left(1-\frac{t_0^2}{2 \rho^2}+\dots \right) - \left(1 + \frac{x_0}{\rho}\right) \left(\frac{t_0}{\rho} - \frac{t_0^3}{6\rho^3}  + \dots \right)} {x_0' \left(\frac{t_0}{\rho} - \frac{t_0^3}{6\rho^3} + \dots \right) +\left(1 + \frac{x_0}{\rho}\right) \left(1-\frac{t_0^2}{2 \rho^2}+\dots \right)} \notag .
\end{equation}
Applying \textbf{Approximation \ref{paraxial approximation}} yields
\begin{equation}
\frac{d \xi}{d \zeta} = \frac{ x_0' - \frac{t_0}{\rho}}{1+\frac{x_0}{\rho}} \notag.
\end{equation}
Now expanding in $x_0 / \rho$, we find that
\begin{equation}
\frac{d \xi}{d \zeta} =\left( x_0' - \frac{t_0}{\rho} \right) \left(1 - \frac{x_0}{\rho} + \frac{x_0^2}{\rho^2} - \dots \right) \notag.
\end{equation}
and, again applying \appr{paraxial approximation},
\begin{equation}
\frac{d \xi}{d \zeta} =x_0' - \frac{t_0}{\rho} \notag.
\end{equation}
Again recalling Eq.~\eqref{xizeta}, we recognize that $\zeta = t$ to first order and, as before, $\xi = x$. Substituting these relations into Eq.~\eqref{xitanbeta}, we have $t= -x \tan \beta$. Finally, substituting $-x_0 \tan \beta$ for $t_0$ into the above equation yields
\begin{equation}
\xi' = x_0' + \frac{x_0}{\rho} \tan \beta. \notag
\end{equation}

To translate the particle through the wedge-shaped drift space, we apply the standard drift space transformation $\xi = \xi_0 + \xi' \zeta_0$ where, to first order, $(x_0,t_0) \equiv (\xi_0, \zeta_0)$ ($\zeta_0$ is the distance from the pole face to the reference plane in the $\zeta$ direction). Rewriting,
\begin{align}
\xi = \xi_0 + \xi' \zeta_0 &= x_0 + \left(x_0' + \frac{x_0}{\rho} \tan \beta\right)t_0 \notag \\
& = x_0 + x_0' t_0 + t_0 \frac{x_0}{\rho} \tan \beta. \notag
\end{align}
Noting from Fig.~\ref{t0pic} that $\frac{t_0}{\rho} \ll 1$ and applying \appr{paraxial approximation}, we eliminate appropriate terms, yielding
$$\xi = x_0.$$
It is important to note that slopes do not change when traversing a drift space, so we still have
$$\xi' = x_0' + \frac{x_0}{\rho} \tan \beta.$$

\begin{figure}[H]
\centering \includegraphics[scale=.9]{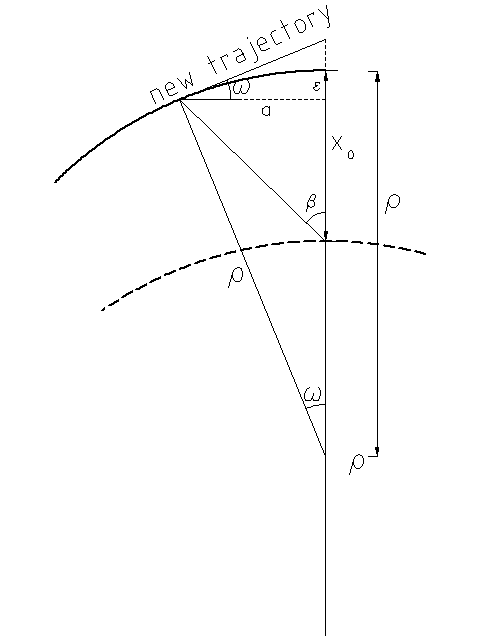} 
\caption{From the triangle with side $a$ opposite of $\beta$, we fnd that $\tan \beta = a / (x_0 - \epsilon)$. From the triangle with $a$ opposite of $\omega$, we have $\tan \omega = a / (\rho -\epsilon)$. By \textbf{Approximation \ref{paraxial approximation}}, we have that $\epsilon \approx 0$. Therefore, $\tan \omega \approx  (x_0 \tan \beta) / \rho$, i.e., the \emph{change} in the slope.}
\label{horiz-trig}
\end{figure}

Then, in order to conform to the standard coordinate system $(x,t)$, we rename $(\xi,\zeta) \equiv (x, t)$ once the particle reaches the reference plane (see Fig.~\ref{coordsys}). Thus we have
\begin{align} \label{sharphoriz}
& x = x_0  \notag \\
& x' = x_0' + \frac{x_0}{\rho} \tan \beta.
\end{align}
This result is the complete transformation due to the fringe field in the horizontal direction. Figure~\ref{horiz-trig} provides an intuitive geometric description of the transformation in the $x'$ coordinate.

\subsection{Vertical Effects}
Vertical effects on rays passing through the fringe field of a dipole are caused by the element of the field perpendicular to the pole face. For the impulse approximation, if $\beta = 0$, this element of the field will be in the direction of motion of the particle, causing no vertical forces on the particle and leaving the trajectory unchanged. To begin analysis, we suppose that the field $\mathbf{B}$ around the pole face also has some non-zero component in the direction perpendicular to the pole face, along with the typical field component $B_y$. It is important to clarify that specifics regarding the field are unnessesary for this mathematical consideration.

We begin with the trajectory coordinates before considering the effects of a vertical fringe field. Note that $y = y_0$ and $y' = y_0'$ before entering the fringe field.

\subsubsection{Equation of Motion}

\begin{figure}[H]
\centering \includegraphics[scale=.35]{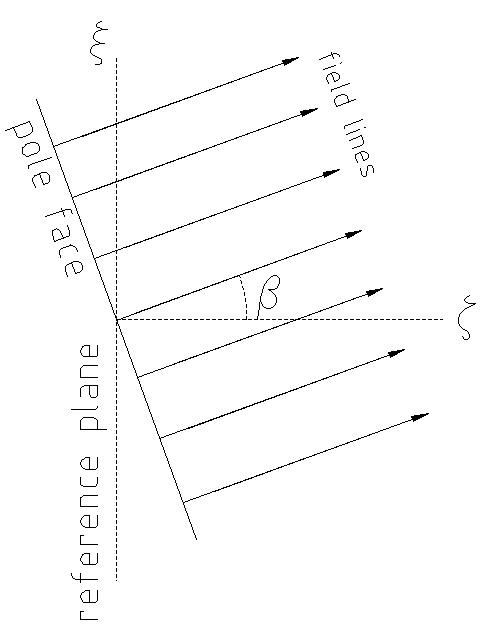} 
\caption{A diagram of the component of $\mathbf{B}$ in the bend plane.}
\label{bcomponents}
\end{figure}

We look to find the equation of motion in the $y$ direction to describe the trajectory. To do so, we must find an expression for all components of the magnetic field. Referencing Maxwell's equations (Ampere's Law), we note that $\mathbf{\nabla} \times \mathbf{B}=\mathbf{0}$. Therefore, 
\begin{align} \label{maxwell}
\frac{\partial B_\zeta}{\partial y} &= \frac{\partial B_y}{\partial \zeta} \notag \\
\frac{\partial B_\xi}{\partial y} & = \frac{\partial B_y}{\partial \xi},
\end{align} 
where the $B_\xi$ and $B_\zeta$ fields are components of the field $\mathbf{B}$ in the respective directions, as seen in Fig.~\ref{bcomponents}.
We recognize that $B_y$ depends solely on $\xi$ and $\zeta$---it is constant in $y$, as expected. Hence, $\frac{\partial B_y}{\partial \zeta}$ depends only on $\xi$ and $\zeta$. With the sharp cutoff field, $B_y$ changes from constant $B_0$ inside of the magnet to zero outside of the magnet. This implies that $B_y$ is a step function where $\frac{\partial B_y}{\partial \zeta}$ is infinite, and therefore nonzero, at the effective field boundary. However, this discontinuity is irrelevant with regards to the following method of analysis.

By using the first line from Eq.~\eqref{maxwell}, $B_\zeta$ is linear in $y$ (upon integrating both sides by $y$), or 
\begin{equation}
B_\zeta = y \frac{\partial B_y}{\partial \zeta}.
\end{equation}

Expressing Newton's Second Law in the $y$ direction in a magnetic field,
\begin{align} \label{newtony}
m \frac{d^2 y}{d t^2} &=q (\mathbf{v} \times \mathbf{B})_y \notag \\
\left(\frac{d^2 y}{d \zeta^2} v_{\zeta} + \frac{dy}{d\zeta} \frac{d v_\zeta}{d\zeta}\right) v_\zeta   & = \frac{q}{m} (\mathbf{v} \times \mathbf{B})_y \notag \\
& = \frac{q}{m} (v_\zeta B_\xi - v_\xi B_\zeta).
\end{align}
where\footnote{Cases involving $v_\zeta=0$, although mathematically feasible, are physically unrealizable and therefore irrelevant.}
\begin{equation}\label{vxi}
v_\xi = \frac{\partial \xi}{\partial \zeta} v_\zeta = \xi' v_\zeta,
\end{equation}
(where the prime notation indicates differentiation with respect to $\zeta$, the longitudinal axis in the $(\xi, y, \zeta)$ coordinate system), from the chain rule. Therefore, $v_\xi B_\zeta = \xi' v_\zeta y \frac{\partial B_y}{\partial \zeta}$. Rewritng Eq.~\eqref{newtony},
\begin{equation}
y'' = \frac{q}{mv_\zeta^2} \left( v_\zeta B_\xi - \xi' v_\zeta y \frac{\partial B_y}{\partial \zeta} - \frac{m}{q} y' \frac{d v_\zeta}{d\zeta} v_\zeta \right) \notag
\end{equation}

\begin{framed}
\begin{approximation} \label{velocity}
The speed $v_0$ of a given ray is
$$v_0 = \sqrt{v_\xi^2 + v_y^2 + v_\zeta^2} = v_\zeta + \frac{v_\xi^2}{2 v_\zeta} + \frac{v_y^2}{2v_\zeta} - \dots .$$
From the chain rule,
$$v_0 = v_\zeta + \frac{\xi'^2 v_\zeta^2}{2 v_\zeta} + \frac{y'^2 v_\zeta^2}{2v_\zeta} - \dots.$$
From \emph{\textbf{Approximation \ref{paraxial approximation}}}, we say
$$v_0 = v_\zeta.$$
\end{approximation}
\end{framed}

By using \textbf{Approximation \ref{velocity}} and the fact that $\rho B_0 = p/q$ (where $B_0$ is the magnitude of the field on the interior of the magnet), we have
\begin{align}
y''&= \frac{q}{mv_0} \left(\frac{v_\zeta B_\xi}{v_\zeta} -  \frac{\xi' v_\zeta y}{v_\zeta} \frac{\partial B_y}{\partial \zeta} - \frac{m}{q} y' \frac{d v_\zeta}{d\zeta} \frac{v_\zeta}{v_\zeta} \right) \notag \\
&=\frac{q}{p}\left( B_\xi -  \xi' y \frac{\partial B_y}{\partial \zeta} - \frac{m}{q} y' \frac{d v_\zeta}{d\zeta} \right) \notag \\
&=\frac{1}{B_0} \left (\frac{B_\xi}{\rho} - \xi' \frac{y}{\rho}  \frac{\partial B_y}{\partial \zeta} - \frac{m}{q \rho} y' \frac{dv_{\zeta}}{d \zeta} \right) \\
&=\frac{1}{B_0} \left (\frac{B_\xi}{\rho} - \xi' \frac{y}{\rho}  \frac{\partial B_y}{\partial \zeta} - \frac{m}{q \rho} y' \frac{d}{dt} \left[ \frac{d\zeta}{d \zeta}\right] \right) \\
&=\frac{1}{B_0} \left (\frac{B_\xi}{\rho} - \xi' \frac{y}{\rho}  \frac{\partial B_y}{\partial \zeta} \right) \notag.
\end{align}
Finally, from \appr{paraxial approximation} in the $(\xi, y, \zeta)$ coordinate system,
\begin{equation} \label{y-diffeq-sc}
y'' = \frac{1}{B_0 \rho} B_\xi.
\end{equation}

\subsubsection{Evaluating the Equation of Motion}
Integrating Eq.~\eqref{y-diffeq-sc} once across the field boundary yields the change in the derivative of $y$ with respect to $\zeta$. That is, 
\begin{equation} \label{deltay'}
\Delta y' (\zeta) = \frac{1}{B_0 \rho} \int_{-\zeta}^{\zeta} B_{\xi}\, d\zeta^*,
\end{equation}
for $\zeta \ge 0$. Referencing Eq.~\eqref{maxwell} and integrating both sides with respect to $\zeta^*$, we find that
\begin{equation} \label{intequiv}
\frac{\partial}{\partial y} \int_{-\zeta}^{\zeta} B_\xi \, d\zeta^* = \frac{\partial}{\partial \xi} \int_{-\zeta}^{\zeta} B_y \, d\zeta^*.
\end{equation}
Recalling Eq.~\eqref{xitanbeta} and differentiating,
\begin{equation} \label{dxidzeta}
\frac{\partial \zeta}{\partial \xi} = -\tan \beta.
\end{equation}
Now consider the integral across the field boundary, $ \int_{-\zeta}^{\zeta} B_y \, d\zeta^*$. Because $B_y$ is $B_0$ before the boundary and zero afterwards, the integral evaluates to simply $B_0 \zeta$. From using this result and by Eq.~\eqref{dxidzeta}, we have
\begin{equation} \label{dxiintegral}
\frac{\partial}{\partial \xi} \int_{-\zeta}^{\zeta} B_y \, d\zeta^* = \frac{\partial}{\partial \xi} (B_0 \zeta) = B_0 \frac{\partial \zeta}{\partial \xi} = -B_0 \tan \beta.
\end{equation}
Now, equating Eq.~\eqref{intequiv} and Eq.~\eqref{dxiintegral} and integrating over $y$ from zero to $y_0$, we have
\begin{equation} \label{almostdelta}
\int_{-\zeta}^{\zeta} B_\xi d \zeta^* = -B_0 y_0 \tan \beta.
\end{equation}
Finally, we can recall Eq.~\eqref{deltay'} and combine it with Eq.~\eqref{almostdelta}, yielding an expression for the change in $y'$,
\begin{equation}
\Delta y' = - \frac{y_0}{\rho} \tan \beta.
\end{equation}

This change in the derivative of $y$ with respect to $\zeta$ is simply the change that results only from crossing the field boundary. We note that an integration to find the change in $y$ would result in infintesimal bounds, yielding a zero result. Hence, the total transformation due to the fringe field is
\begin{align} 
& y = y_0 \notag \\
& y' = y_0' - \frac{y_0}{\rho} \tan \beta.
\end{align}

\subsection{Alternative Derivation of Vertical Effects}
\begin{figure}[H]
\centering \includegraphics[scale=.4]{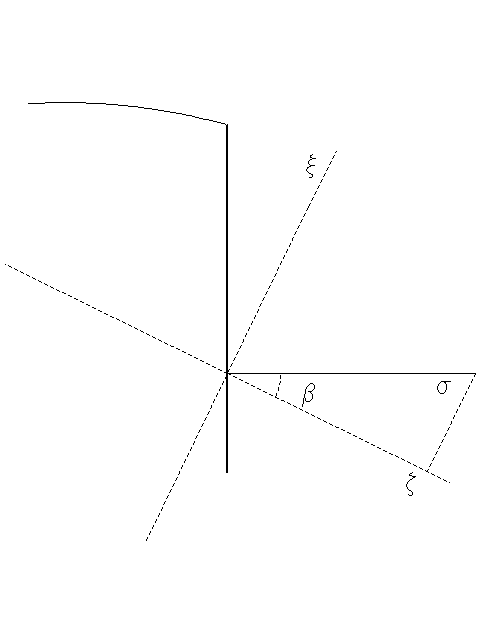} 
\caption{Graphical depiction of the geometry used to define $\sigma$.}
\label{sigma-pic}
\end{figure}

We begin by considering the vertical component of the magnetic field in the magnet. On the interior of the magnet, the vertical component of the field has a value of $B_0$ and on the exterior of the magnet the vertical component of the field has a value of zero. We look to define a new axis that is perpendicular to the rotated pole-face of the magnet in order to express the field mathematically. We use the geometry of the $\xi$ and $\zeta$ axes and the pole face to construct this axis (see Fig.~\ref{sigma-pic}). Let us define the new axis $\sigma$ as 
$$\sigma = \xi \sin \beta + \zeta \cos \beta.$$
\noindent The field along the $\sigma$ axis looks like a Heaviside function, $H(x)$, so
$$B_y = B_0 \,H(-\sigma)$$
Rewriting this definition in terms of $\xi$ and $\zeta$ gives
$$B_y = B_0 \, H (-\xi \sin \beta - \zeta \cos \beta).$$

In order to find the field components in the $\xi$ and $\zeta$ directions, we must reference one of Maxwell's equations and see that $\nabla \times \mathbf{B} = \mathbf{0}$. That is,
\begin{equation}
\left| \begin{array}{ccc}
{\hat \xi} & {\hat y} & {\hat \zeta} \\
\frac{\partial}{\partial \xi} & \frac{\partial}{\partial y} & \frac{\partial}{\partial \zeta} \\
B_\xi & B_y & B_\zeta \end{array} \right| = \mathbf{0}. \notag
\end{equation}
Hence, we have
\begin{align}
 \frac{\partial B_\xi}{\partial y} &= \frac{\partial B_y}{\partial \xi} = - B_0 \delta (\xi \sin \beta + \zeta \cos \beta)  \sin \beta \notag \\
 \frac{\partial B_\zeta}{\partial y} &= \frac{\partial B_y}{\partial \zeta} = - B_0 \delta (\xi \sin \beta + \zeta \cos \beta)  \cos \beta, \notag
\end{align}
where $\delta (x)$ is the Dirac delta function.

Therefore, we can solve for $B_\xi$ and $B_\zeta$ by integrating from zero to $y_0$ in $y$. Because these functions are constant in $y$, this calculation is a simple process and yields, together with $B_y$,
\begin{align}\label{bvector}
B_\xi &= - B_0y_0 \delta (\xi \sin \beta + \zeta \cos \beta)  \cos \beta \notag \\
B_y &= B_0 \, H (-\xi \sin \beta - \zeta \cos \beta) \notag \\
B_\zeta &= - B_0y_0 \delta (\xi \sin \beta + \zeta \cos \beta) \sin \beta \notag.
\end{align}

As before, we write the equation of motion (see Eq.~\eqref{newtony}) and substitute in the field components and the expression for $v_\xi$ (as in Eq.~\eqref{vxi}). So, we now have
\begin{align}
\left(y'' v_{\zeta} + y' \frac{d v_\zeta}{d\zeta}\right) v_\zeta & = \frac{q}{m} (\mathbf{v} \times \mathbf{B})_y \notag \\
& = \frac{q}{m} (v_\zeta B_\xi - v_\xi B_\zeta) \notag \\
& = \frac{q B_0 v_\zeta}{m} \left [\xi ' y_0 \cos \beta - y_0 \sin \beta \right] \,\delta(\xi \sin \beta + \zeta \cos \beta) \notag \\
y'' & = \frac{1}{\rho} \left [\xi ' y_0 \cos \beta - y_0 \sin \beta - \frac{m}{q} y' \frac{d}{dt} \left(\frac{d\zeta}{d\zeta} \right) \right] \, \delta(\xi \sin \beta + \zeta \cos \beta) \\
y'' & = \frac{1}{\rho} \left [\xi ' y_0 \cos \beta - y_0 \sin \beta \right] \, \delta(\xi \sin \beta + \zeta \cos \beta) \notag.
\end{align}
By \appr{paraxial approximation}, we eliminate the second order term $\xi' y/\rho$. We simplify the equation to
\begin{equation}\label{eom-y2-simp}
y''  = -\frac{y_0}{\rho} \delta(\xi \sin \beta + \zeta \cos \beta) \sin \beta.
\end{equation}
Integrating with respect to $\zeta$ across the pole-face from $-\xi \tan \beta -\epsilon$ to $-\xi \tan \beta + \epsilon$ (referencing Eq.~\eqref{xitanbeta}) for any $\epsilon > 0$, we have
\begin{align}
\Delta y' &= -\frac{y_0 \sin \beta}{\rho} \int_{-\xi \tan (\beta) -\epsilon}^{-\xi \tan (\beta) + \epsilon} \delta(\xi \sin \beta + \zeta \cos \beta) \, d \zeta \notag \\
\Delta y' &= -\frac{y_0 \tan \beta}{\rho} \notag.
\end{align}
This result is the total change of the slope $y'$ from passing through the sharp cutoff field.

We then attempt to calculate the total change in the $y$ position from passing through the pole face. To do this, we take the antiderivative of both sides of Eq.~\eqref{eom-y2-simp} and integrate across the field boundary, as before, with the same bounds.
\begin{align}
\int_{-\epsilon}^{\epsilon} \frac{dy}{d\zeta} d\zeta &= \int_{-\xi \tan (\beta) -\epsilon}^{-\xi \tan (\beta)+\epsilon} \left[- \frac{y_0 \sin \beta}{\rho} \int \delta(\xi \sin \beta + \zeta^* \cos \beta) \, d \zeta^* \right] \, d \zeta \notag \\
\notag \\
\Delta y &=  \int_{-\xi \tan (\beta) -\epsilon}^{-\xi \tan (\beta) +\epsilon} \left[c - \frac{y_0 \tan \beta \, H(\xi \sin \beta + \zeta \cos \beta)}{\rho}\right] \, d \zeta \notag \\
\notag \\
\Delta y &= \left[c \zeta - \frac{y_0 \tan \beta}{\rho} H(\xi \sin \beta + \zeta \cos \beta) (\zeta + \xi \sin \beta) \right] \bigg |_{-\xi \tan (\beta) -\epsilon}^{-\xi \tan (\beta) +\epsilon} \notag
\end{align}
We see as $\epsilon$ goes to zero,
\begin{equation}
\Delta y = 0. \notag
\end{equation}

The above result shows that the total change of $y$ from passing through the sharp cutoff field is zero.

Thus, the transformation through the sharp cutoff field is:
\begin{align}
y&= y_0  \notag\\
y' &= y_0' -\frac{y_0 \tan \beta}{\rho} \notag.
\end{align}

\section{Extended Fringe Field Effects}
\subsection{Coordinate System}
\begin{figure}[H]
\centering \includegraphics[scale=.3]{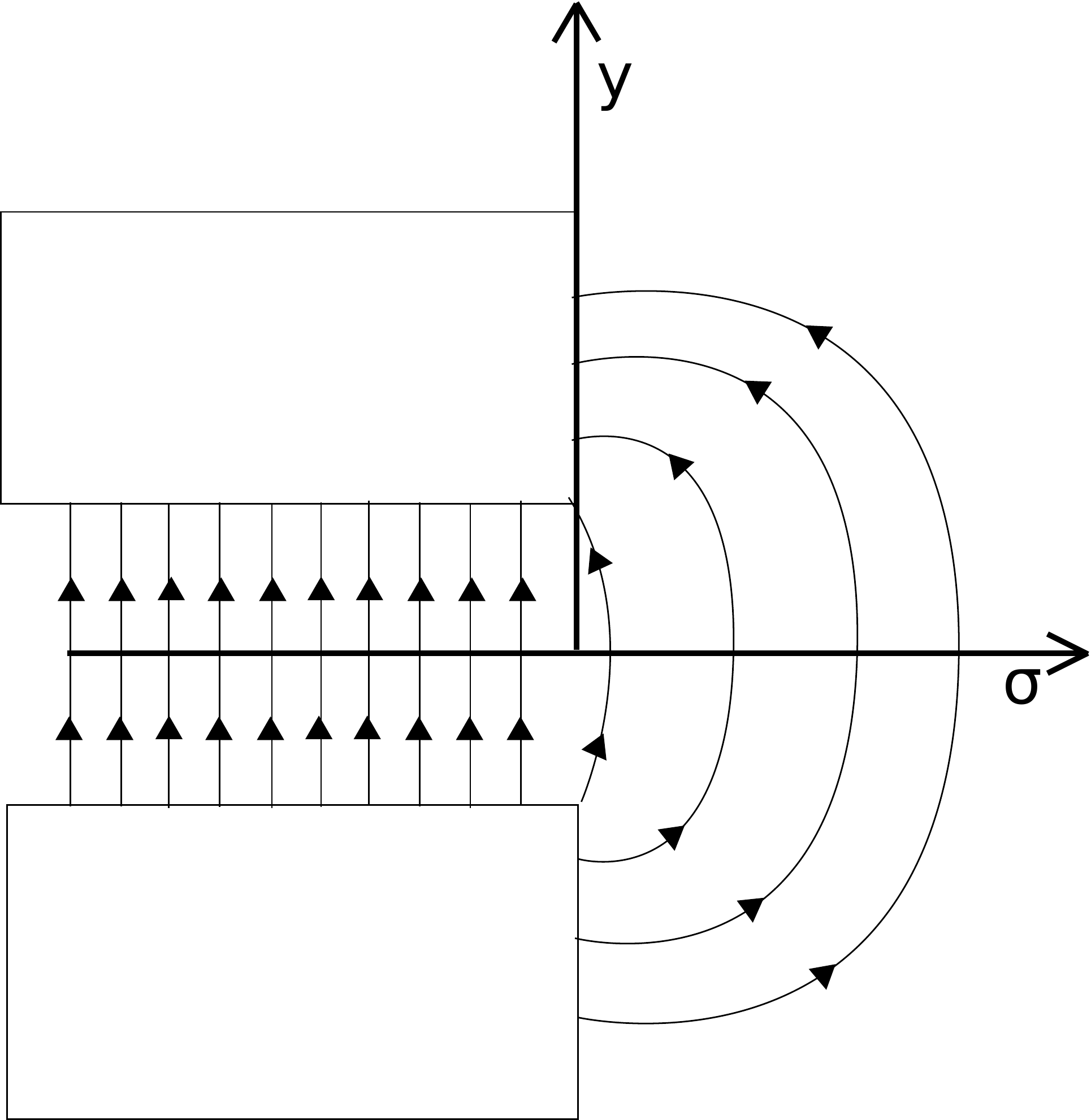} 
\caption{A view from the side of the dipole exit face. There is no field in or out of the page.}
\label{fieldgraphic}
\end{figure}

In this section, we look to get a better approximation than the previous impulse approximation. The impulse approximation's sharp cutoff magnetic field cannot be physically realized. We look to find an approximation based on a finite vertical pole gap where the fringe field extends beyond that of the sharp cutoff case. Recall that the pole width is still treated as if it were infinite so that the field never has a non-zero horizontally transverse component. See Fig.~\ref{fieldgraphic}. 

We first look to define an axis perpendicular to the pole-face in the $y=0$ plane (see Fig.~\ref{sigma-pic}). The magnitude of the field $|\mathbf{B}|$ is then only a function of this coordinate $\sigma$ and the $y$ coordinate. That is, 
\begin{equation} \label{sigmadef}
\sigma = \frac{\zeta \cos \beta + \xi \sin \beta}{g},
\end{equation}
where $g$ is the vertical pole gap\footnote{The definition of $\sigma$ here differs from the definition given in the alternative derivation for the sharp cutoff field by a factor of $g$.}. For sufficiently large positive $\sigma$, the magnitude of the field is zero. Conversely, for sufficiently large negative $\sigma$, the magnitude of the field is $B_0$, or the magnitude inside the magnet. We name these values $\sigma_2$ and $-\sigma_1$ respectively. Note that $\sigma_2$ is finite but large enough such that any remaining field is trivial. 

\begin{figure}[H]
\centering \includegraphics[scale=.4]{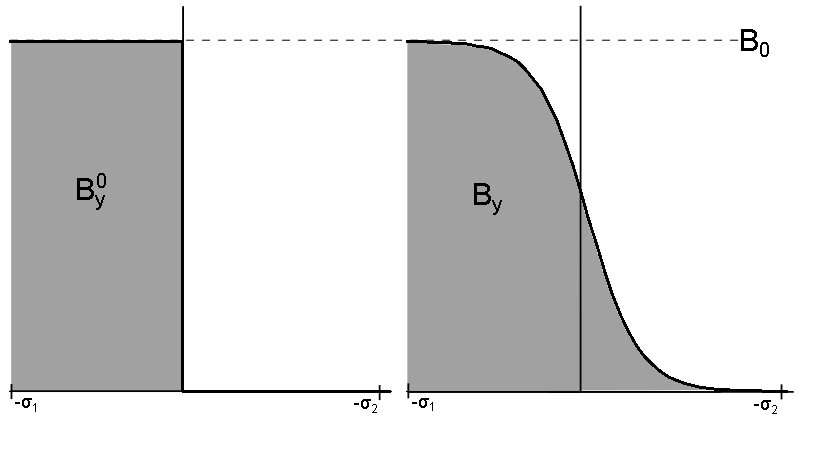} 
\caption{A graphical depiction of Eq.~\eqref{intequal}. The gray regions are equivalent in area, showing the equivalence of the integrals.}
\label{fieldintegral}
\end{figure}

Let $\sigma = 0$ be the ``effective field boundary," or the point at which the sharp cutoff field drops to zero. Further, let $B^0_y$ be the function representing the sharp cutoff field. Consider the sharp cutoff field and the extended fringe field as functions of $\sigma$. We set the effective field boundary such that 
\begin{equation} \label{intequal}
\int_{-\sigma_1}^{\sigma_2} B_y \, d \sigma = \int_{-\sigma_1}^{\sigma_2} B_y^0 \, d \sigma = \sigma_1 B_0,
\end{equation}
where $B_y$ represents the magnitude of the field in the vertical direction at $y=0$ (see Fig.~\ref{fieldintegral}). While the definition is an approximation because $B_y$ does indeed vary with $y$, it is necessary and reasonable for small, practical values of $y$ (as can be seen in Fig.~\ref{fieldgraphic}).

Finding the field components in the other two directions $\xi$ and $\zeta$ is necessary in order to apply the Lorentz force. Because $B_y = B_y(\sigma)$, we know that $B_y = B_y(\xi,\zeta)$. Applying Ampere's Law,
$$\frac{\partial B_\zeta}{\partial y} = \frac{\partial B_y}{\partial \zeta}, \, \, \, \frac{\partial B_\xi}{\partial y} = \frac{\partial B_y}{\partial \xi}.$$
Taking the antiderivative of both sides of each of the above equations with respect to $y$ yields
\begin{align}
B_\zeta = \int \frac{\partial B_y}{\partial \zeta} dy \notag \\
B_\xi = \int \frac{\partial B_y}{\partial \xi} dy \notag
\end{align}
Evaluating the integrals on the right, we have
\begin{align}
B_\zeta =  y \frac{\partial B_y}{\partial \zeta} \notag \\
B_\xi =  y \frac{\partial B_y}{\partial \xi} \notag
\end{align}
Finally, applying the chain rule from derivatives of Eq.~\eqref{sigmadef},
\begin{align} \label{bcomp}
& B_\zeta = \frac{y \cos \beta}{g} \frac{dB_y}{d\sigma}\\
& B_\xi = \frac{y \sin \beta}{g} \frac{dB_y}{d\sigma} \notag.
\end{align}
From these expressions of the field in the three coordinate directions, we can determine the horizontal and vertical effects of the fringe field.

\subsection{Horizontal Effects}
To calculate the horizontal trajectory, we begin with an expression of Newton's second law in the $\xi$ direction,
\begin{align}
m\left(\xi'' v_\zeta + \xi' \frac{d v_\zeta}{d \zeta}\right)v_\zeta & = q (\mathbf{v} \times \mathbf{B})_{\xi} \notag \\
& = q ( v_y B_\zeta - v_\zeta B_y) \notag
\end{align}
From Eq.~\eqref{bcomp}, we substitute in $B_\zeta$. Further we recognize that $v_y = v_\zeta y'$ where the prime indicates differentiation with respect to $\zeta$ (the longitudinal axis in this coordinate system). So,
\begin{equation}
m\left(\xi'' v_\zeta + \xi' \frac{d v_\zeta}{d \zeta}\right)v_\zeta = q \left ( v_\zeta y' y \frac{\cos \beta}{g} \frac{d B_y}{d \sigma} - v_\zeta B_y \right) \notag
\end{equation}
After dividing by $m v_\zeta$, we use \appr{velocity} and $p/q = B_0 \rho$:
\begin{align}
\xi'' v_\zeta + \xi' \frac{d v_\zeta}{d \zeta} &= \frac{q}{m} \left( y' y \frac{\cos \beta}{g} \frac{d B_y}{d \sigma} -  B_y \right) \notag \\
\xi'' v_\zeta &= \frac{q}{m} \left( y' y \frac{\cos \beta}{g} \frac{d B_y}{d \sigma} -  B_y - \frac{m}{q} \xi' \frac{d}{dt} \left[\frac{d \zeta}{d \zeta}\right] \right) \notag \\
\xi''&= \frac{1}{B_0 \rho} \left( y' y \frac{\cos \beta}{g} \frac{d B_y}{d \sigma} -  B_y \right). \notag
\end{align}
Then, applying \appr{paraxial approximation}, we eliminate the term containing $y' y/\rho$ and find
\begin{align} \label{xi''}
\xi'' = - \frac{1}{B_0 \rho} B_y.
\end{align}
Because $ d \zeta/d\sigma = g/\cos \beta$ from Eq.~\eqref{sigmadef}, we multiply both sides of Eq.~\eqref{xi''} by this quantity and apply the chain rule to both sides, yielding
\begin{equation} \label{dxids}
\frac{d}{d\sigma} [\xi']  = - \frac{g}{B_0 \rho \cos \beta} B_y.
\end{equation}
Integrating both sides from $-\sigma_1$ to $\sigma_2$,
\begin{equation} \label{diffxi'}
\xi' \big|_{\sigma_2} - \xi'\big|_{-\sigma_1} = - \frac{g}{B_0 \rho \cos \beta} \int_{-\sigma_1}^{\sigma_2} B_y d \sigma 
\end{equation}
We observe that Eq.~\eqref{diffxi'} is valid for $B^0_y$ and $B_y$, the extended field case. We denote the slopes of the trajectories with subscripts $E$ and $S$ for the extended fringe field and sharp cutoff field cases respectively. Subtracting the two, we have
\begin{equation} 
\left[ \xi'_S \big|_{\sigma_2} - \xi'_S \big|_{-\sigma_1} \right] - \left[ \xi'_E \big|_{\sigma_2} - \xi'_E \big|_{-\sigma_1} \right] = - \frac{g}{B_0 \rho \cos \beta} \int_{-\sigma_1}^{\sigma_2} (B^0_y - B_y) \, d \sigma. \notag
\end{equation}
Recognizing that the second and fourth terms on the left side both represent the slope well inside the magnet by the definition of $\sigma_1$, their difference is zero. Thus, 
\begin{equation}
\xi'_S \big|_{\sigma_2} - \xi'_E \big|_{\sigma_2} = - \frac{g}{B_0 \rho \cos \beta} \int_{-\sigma_1}^{\sigma_2} (B^0_y - B_y) \, d \sigma \notag.
\end{equation}
This equation represents the difference of the slopes of the sharp cutoff case and the extended case at $\sigma_2$. Recalling Eq.~\eqref{intequal},
$$ \int_{-\sigma_1}^{\sigma_2} (B^0_y - B_y) \, d \sigma = (B_0 \sigma_1 - B_0 \sigma_1)=0.$$ 
Then,
$$\xi'_S \big|_{\sigma_2} - \xi'_E \big|_{\sigma_2} = 0.$$
Hence, the vertical slope of the trajectory in the bend plane in the case of the extended fringe field is equal to the slope in the sharp cutoff field. As such, $x' = x_0' + \frac{x_0}{\rho} \tan \beta$, just as in Eq.~\eqref{sharphoriz}.

By using the process to achieve Eq.~\eqref{dxids} twice, we find that
$$\frac{d^2}{d \sigma^2} [\xi] = -\frac{g^2}{B_0 \rho \cos^2 \beta} B_y.$$
We integrate twice with respect to $\sigma$ and subtract the sharp cutoff case and the extended case to find the difference in $\xi$ between the extended case and the sharp cutoff case, just as before. This process results in
\begin{align}
\xi_E \big|_{\sigma_2} - \xi_S \big|_{\sigma_2} &=  \frac{g^2}{B_0 \rho \cos^2 \beta} \int_{-\sigma_1}^{\sigma_2} \int_{-\sigma_1}^{\sigma^*} (B^0_y - B_y ) \, d \sigma \, d \sigma^* \notag \\
 &= \frac{g^2}{\rho \cos^2 \beta } I_1 \notag
\end{align}
where
\begin{equation}
I_1 = \int_{-\sigma_1}^{\sigma_2} \int_{-\sigma_1}^{\sigma^*} \frac{B^0_y - B_y}{B_0} \, d \sigma \, d \sigma^*. \notag
\end{equation}
This difference represents the horizontal dispacement of a ray from the equivalent ray in the sharp cutoff case. This displacement affects any input ray equivalently. Therefore, the net effect is a shift in the entire beam, including the reference trajectory. Because this difference does not depend on any initial trajectory coordinates, we call this effect a ``zeroth order" effect.

The final effects of the fringe field in the horizontal direction are
\begin{align}
x = x_0 + \frac{g^2}{\rho \cos^2 \beta } I_1 \notag \\
x' = x_0' + \frac{x_0}{\rho} \tan \beta.
\end{align}

\subsection{Vertical Effects}

\begin{figure}[H]
\centering \includegraphics[scale=.7]{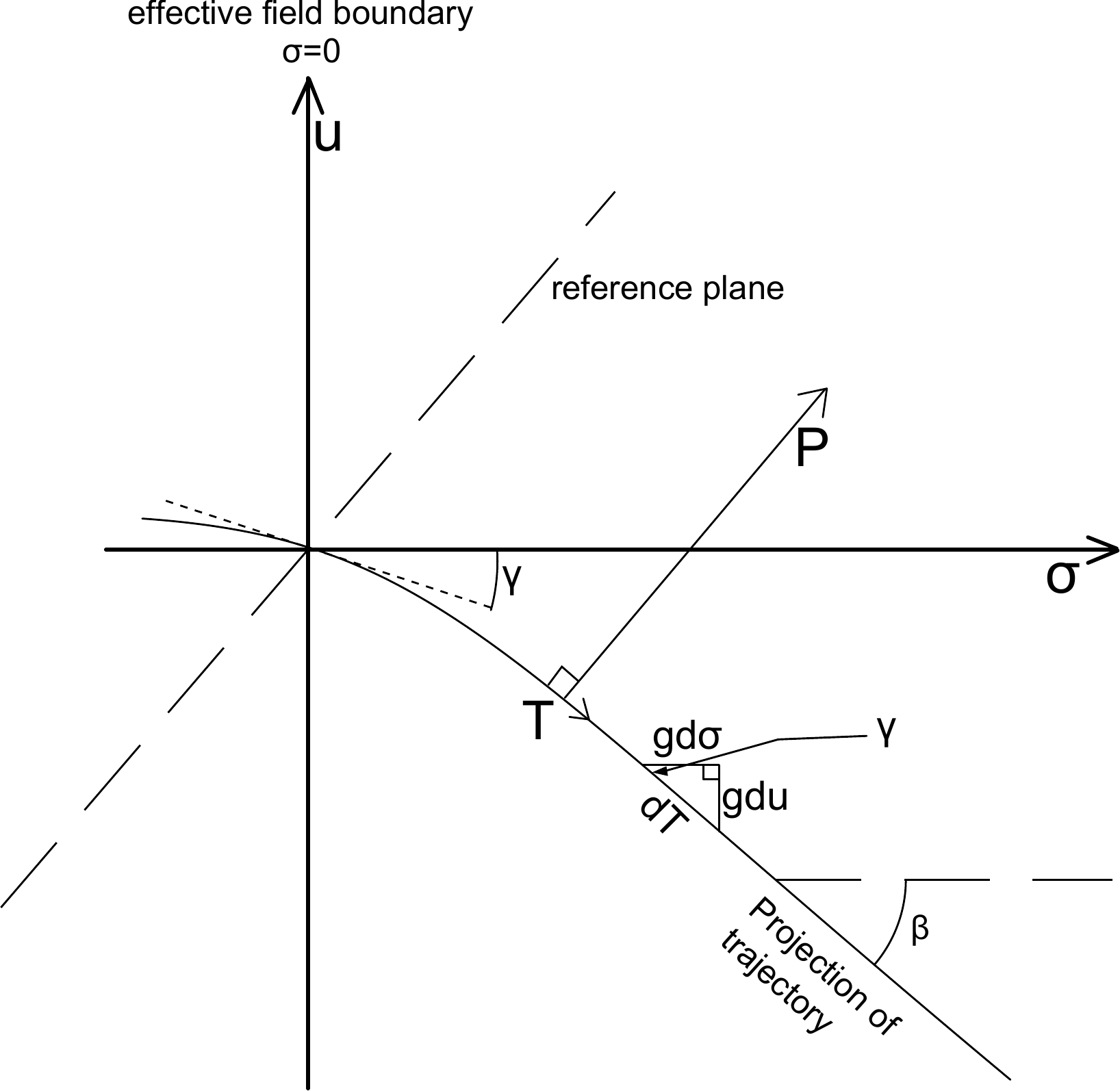} 
\caption{A graphical description of the coordinate systems used in this section, with the $y$ axis pointing out of the page. Note that the $T$ and $P$ axes follow the trajectory of each particle (in this case, the particle's trajectory in the plane is parallel to the reference trajectory). The angle $\gamma$ lies between the $\sigma$ axis and a line tangent to the trajectory.}
\label{gamma-pic}
\end{figure}

To adequately consider the vertical effects, we must consider the horizontal trajectory throughout the fringe field and how it affects the vertical motion. It is important to recognize that, due to the fringe field, not even the reference particle exits the magnet parallel to the $\zeta$ axis: because the fringe field is continuously decreasing to zero as opposed to a instantaneous change in the sharp cutoff case, both particles experience lesser horizontal bending effects but for longer periods of time. This phenomenon implies that particles are at an angle to the $\sigma$ axis, called $\gamma$, which continues to change until 
the particles are a significant distance from the pole face, at which point $\gamma(\sigma_2) = \beta$. See Fig.~\ref{gamma-pic}.

Suppose a particle displaced in the $y$ direction. Because of the particle's position along the $y$ axis, it will experience a field pointing in the $\sigma$ direction as well as the typical field in the $y$ direction while within the fringe field (see Fig.~\ref{fieldgraphic}). The field component in the direction perpendicular to the trajectory within the bend plane is the direct cause of the vertical focusing/defocusing effects on the displaced particle. Thus, the vertical effects depend on the said field component and the speed of the particle projected into the bend plane, $v_T$. Let $(P,y,T)$ be a curvilinear coordinate system where the $P$ axis is always perpendicular to the particle's trajectory $T$ in the bend plane.

\subsubsection{Equation of Motion}
By using the same technique that we used to find the field components in Eq.~\eqref{bcomp}, we can find $B_P$ and $B_T$:
\begin{align} \label{BtBp}
B_T = y \frac{\cos \gamma}{g} \frac{d B_y}{d \sigma} \notag  \\
B_P = y \frac{\sin \gamma}{g} \frac{d B_y}{d \sigma}.
\end{align}
We then consider the equation of motion in the vertical direction,
\begin{equation}
\left(\frac{d^2y}{dT^2} v_T + \frac{dy}{dT} \frac{dv_T}{dT} \right) v_T = v_T^2 \frac{d^2 y}{dT^2}= \frac{q}{m} (v_T B_P - v_P B_T). \notag
\end{equation}
Because of the definition of the coordinate system, the only velocity in the bend plane is tangential to the trajectory, i.e. $v_P=0$. Further, applying \appr{velocity} in the $(P,y,T)$ coordinate system, $v_T = v_0$. Simplifying, 
\begin{equation} 
\frac{d^2 y }{d T^2} = \frac{1}{B_y \rho} B_P \notag
\end{equation}
Then, from Eq.~\eqref{BtBp}, 
\begin{equation} \label{motionverticalext}
y'' = \frac{1}{B_0 \rho} y \frac{\sin \gamma}{g} \frac{d B_y}{d \sigma}, 
\end{equation}
where prime indicates differentiation with respect to $T$ (again the longitudinal direction in this coordinate system).

Consider the differential right triangle in Fig.~\ref{gamma-pic} (we include $g$ because the distance along the $\sigma$ axis must be scaled to the distance along the $T$ axis). Trigonometric relationships yield
\begin{equation}\label{dtds}
\frac{dT}{d\sigma} = \frac{g}{\cos \gamma}.
\end{equation}
We multiply the left side of Eq.~\eqref{motionverticalext} by $dT$ and the right side by $\frac{g}{\cos \gamma} \, d\sigma$. Then, we integrate from $\sigma = \sigma^*$  to $ \sigma_2$, where $\sigma^*$ is a floating variable.
\begin{align} \label{dydT}
 \int_{T(\sigma^*)}^{T(\sigma_2)} \frac{d^2y}{dT^2} \, dT &= \frac{1}{B_0 \rho} \int_{\sigma^*}^{\sigma_2} y \tan \gamma \frac{dB_y}{d\sigma} \, d\sigma \notag \\
\frac{dy}{dT} \bigg|_{\sigma_2} - \frac{dy}{dT} \bigg|_{\sigma^*} &= \frac{1}{B_0 \rho} \int_{\sigma^*}^{\sigma_2} y \tan \gamma \frac{dB_y}{d\sigma} \, d\sigma \\
\Delta \frac{dy}{dT} &= \frac{1}{B_0 \rho} \int_{\sigma^*}^{\sigma_2} y \tan \gamma \frac{dB_y}{d\sigma} \, d\sigma \notag
\end{align}

Now, we must find approximations for $y$ and $\tan \gamma$. In the end, we will substitute these back into the previous expression, yielding the total change in $y'$. 
\subsubsection{Expression for $\gamma$}
We now attempt to find an expression containing $\gamma$ from a differential equation of motion in the perpendicular direction from $d \gamma/d T$. 

 We begin expressing the Lorentz force in the perpendicular direction in terms of $\gamma$ and $T$. By linking $T$ and $\gamma$ with the instantaneous radius of curvature $R \equiv R(T)$ of the path, we recognize that $dT = R \, d\gamma$. Utilizing $R$ in the equation for centripetal acceleration and equating it to the Lorentz force, we have
\begin{align} 
-m v_T^2 \frac{1}{R} = -m v_T^2 \frac{d \gamma}{d T} &= q (v_y B_T - v_T B_y) \notag \\
&= q \left(y'v_T y \frac{\cos \gamma}{g} \frac{d B_y}{d \sigma} - v_T B_y \right). \notag
\end{align}
As used several times previously, we apply \appr{paraxial approximation} and \appr{velocity} but in the $(P,y,T)$ coordinate system. Hence, the equation of motion in the perpendicular direction is
\begin{equation} \label{dgdt}
\frac{d \gamma}{d T} = \frac{1}{B_0 \rho} B_y.
\end{equation}
Recall from Eq.~\eqref{dtds} that $dT = \frac{g}{\cos \gamma} \, d\sigma$. Applying this relation to Eq.~\eqref{dgdt}, we have that
\begin{equation} \label{eomsec}
\frac{d \gamma}{d \sigma} = \frac{gB_y}{B_0 \rho} \sec \gamma.
\end{equation}

Next, suppose $u$ is the axis perpendicular to the $\sigma$ axis (see \cite{theis}), scaled in $g$ (just as $\sigma$ is). As before, we consider the differential right triangle in Fig.~\ref{gamma-pic}. We see
\begin{equation}
\tan \gamma = \frac{du}{d\sigma}. \notag
\end{equation}
Differentiating with respect to $\sigma$,
\begin{equation}
\frac{d^2 u}{d \sigma ^2} = \frac{d \gamma}{d \sigma} \sec ^2 \gamma. \notag
\end{equation}
Substituting Eq.~\eqref{eomsec} into the previous equation, we have
\begin{equation} \label{final-gamma-motion}
\frac{d^2 u}{d \sigma^2} = \frac{g B_y}{B_0 \rho} \sec^3 \gamma.
\end{equation}

We integrate\footnote{Note that $\sigma^*$ is simply a dummy variable used to clarify the discrepancy between the variable of integration and the bounds.} both sides of Eq.~\eqref{final-gamma-motion} from $\sigma$ to $\sigma_2$, where $\sigma$ is a floating variable,
\begin{align}
\frac{du}{d\sigma} \big|_{\sigma_2} - \frac{du}{d\sigma} \big |_{\sigma} &= \int_{\sigma}^{\sigma_2} \frac{g B_y}{B_0 \rho} \sec^3 \gamma \, d \sigma^* \notag \\
\tan (\gamma (\sigma_2)) - \tan (\gamma (\sigma)) &=  \frac{g}{B_0 \rho} \int_{\sigma}^{\sigma_2} B_y \sec^3 \gamma \, d \sigma^*. \notag
\end{align}
Substituting in $\gamma(\sigma_2) = \beta$ and that $\gamma(\sigma) \equiv \gamma$,
\begin{align}
 \tan \gamma &= \tan \beta - \frac{g}{B_0 \rho} \int_{\sigma}^{\sigma_2} B_y \sec^3 \gamma \, d \sigma^*
\end{align}
By using an iterative approximation\footnote{By iterative approximation, we mean a process of utilizing some given information and inputting it to one side of some implicit equation to approximate all other values described in said equation. In this case, we use fact that $\gamma (\sigma_2) = \beta$ and substitute this value into the right-hand side of the equation.}, utilizing the given point of $\gamma (\sigma_2) = \beta$, we find that
\begin{equation} \label{gamma-iterative}
\tan \gamma = \tan \beta - \frac{g}{B_0 \rho \cos^3 \beta} \int_{\sigma}^{\sigma_2} B_y \, d \sigma^*.
\end{equation}

\subsubsection{Expression for $y$}


We look to use a similar iterative approximation to find an equation for $y(\sigma)$ in terms of some fixed point $y_2$ by using Eq.~\eqref{dydT}. 

Rewriting Eq.~\eqref{dydT}, we have
\begin{equation}
\frac{dy}{dT} = y_2'-\frac{1}{B_0 \rho} \int_{\sigma^*}^{\sigma_2} y \tan \gamma \frac{dB_y}{d \sigma} \, d \sigma, \notag
\end{equation}
where $y_2'= \frac{d y}{dT} \big|_{\sigma_2}$. The next step is to make the iterative approximation. We choose the fixed point as $\gamma(\sigma_2) = \beta$ and $y(\sigma_2) = y_2$. Simplifying,
\begin{align} \label{yconst}
\frac{dy}{dT} &= y_2' -\frac{y_2 \tan \beta}{B_0 \rho} \int_{\sigma^*}^{\sigma_2} \frac{dB_y}{d \sigma} \, d \sigma \notag \\
&=y_2' + \frac{y_2 \tan \beta}{B_0 \rho} B_y (\sigma^*).
\end{align}
Then, integrating from $\sigma^* = \sigma$ to $\sigma_2$ and solving for $y$, we have
\begin{align}
y_2 - y(\sigma)& = \frac{y_2' g(\sigma_2 - \sigma)}{\cos \beta} + \frac{y_2 g \tan \beta}{B_0 \rho \cos \beta} \int_{\sigma}^{\sigma_2} B_y \, d \sigma^*. \notag
\end{align}
where the $\nicefrac{g}{\cos \beta}$ factor above arises from applying the chain rule to convert from $dT$ to $d \sigma^*$.

\begin{framed} 
\begin{approximation}\label{thin lens}
Physically, we know that the extent of the fringe field, $g(\sigma_2+\sigma_1)$, is of the same order as the vertical gap $g$. Because $g/L \ll 1$ where $L$ is the length of the magnet, we conclude that 
$$\frac{g (\sigma_2 + \sigma_1)}{L}\ll 1.$$
\end{approximation}
\end{framed}

We can see from division by $L$ that the term containing $y_2' g(\sigma_2-\sigma)$ contains the product of two small values (assuming $-\sigma_1 \le \sigma \le \sigma_2$) and can be neglected, following \appr{thin lens}. Then simplifying, 
\begin{align} \label{ysigma}
y(\sigma) &= y_2 \left( 1 - \frac{g \tan \beta}{B_0 \rho \cos \beta} \int_{\sigma}^{\sigma_2} B_y \, d \sigma^* \right).
\end{align}

\subsubsection{Change in $y'$}
Utilizing the previous three subsections, we derive an approximation for the change in $y'$, the vertical slope, as a function of the vertical displacement, $y_2$. Substituting in Eq.~\eqref{gamma-iterative} and Eq.~\eqref{ysigma} into Eq.~\eqref{dydT}, we have
\footnotesize
\begin{align}
\begin{split}
&\Delta \frac{dy}{dT} = \frac{1}{B_0 \rho} \int_{\sigma^*}^{\sigma_2} y \tan 
\gamma \frac{dB_y}{d\sigma} \, d\sigma \notag \\
& = \frac{1}{B_0 \rho} \int_{\sigma^*}^{\sigma_2} y_2 \left( 1 - \frac{g \tan \beta}{B_0 \rho \cos \beta} \int_{\sigma}^{\sigma_2} B_y \, d \sigma^* \right) \left(\tan \beta \notag -\frac{g}{B_0 \rho \cos^3 \beta} \int_{\sigma}^{\sigma_2} B_y \, d\sigma^* \right) \frac{dB_y}{d\sigma} \, d\sigma \notag \\
& = -\frac{y_2}{B_0 \rho} \int_{-\sigma_*}^{\sigma_2} \bigg( \tan \beta - \bigg[ \frac{g}{B_0 \rho \cos^3 \beta} \int_{\sigma}^{\sigma_2} B_y d \sigma^* \bigg] - \bigg[ \frac{g \sin^2 \beta}{B_0 \rho \cos^3 \beta} \int_{\sigma}^{\sigma_2} B_y d \sigma^* \bigg] \notag \\
&\, \; \; \, \; + \frac{g^2 \tan \beta}{B_0^2 \rho^2 \cos^4 \beta } \bigg( \int_{\sigma}^{\sigma_2} B_y d \sigma^* \bigg)^2 \bigg) \frac{d B_y}{d\sigma} d\sigma.
\end{split}
\end{align}
\normalsize
The lower bound of the outer integral $\sigma^*$ is set to $-\sigma_1$ to account for the effects of the entire fringe field.

\begin{framed}
\begin{approximation} \label{gap size}
For a length of a magnet $L$ on the order of or smaller than the radius of curvature $\rho$ and assuming $g/L \ll 1$ as in \appr{thin lens}, it follows that 
$$\frac{g}{\rho} \ll 1.$$ 
\end{approximation}
\end{framed}

Eliminating all but first order terms in $g/\rho$ from \appr{gap size} and simplifying, 
\begin{equation} \label{finaldy}
\Delta \frac{dy}{dT} = - \frac{y_2}{\rho} \left( \tan \beta + g \left( \frac{1+ \sin^2 \beta}{B_0^2 \rho \cos^3 \beta} \right) \int_{-\sigma_1}^{\sigma_2} \left[ \frac{d B_y}{d \sigma} \int_{\sigma}^{\sigma_2} B_y d \sigma^* \right] d \sigma \right). 
\end{equation}

We can further simplify Eq.~\eqref{finaldy} with integration by parts to see that 
\begin{equation}
\int_{-\sigma_1}^{\sigma_2} \left[ \frac{d B_y}{d \sigma} \int_{\sigma}^{\sigma_2} B_y d \sigma^* \right] d \sigma = - \int_{-\sigma_1}^{\sigma_2} B_y ( B_0 - B_y) d \sigma
\end{equation}

To compare this result with the sharp cutoff case, suppose an ``effective exit angle" $\beta_v$ such that
\begin{equation}
\tan \beta_v = \tan \beta - \frac{g}{\rho} \frac{1 + \sin^2 \beta}{\cos^3 \beta} I_2,
\end{equation}
where $I_2$ is a dimensionless, measurable integral\footnote{Note that within texts like \emph{The Optics of Charged Particle Beams}, this integral contains $g$ in the denominator; however, a different definition of $\sigma$ lacking $g$ is used. The definition Eq.~\eqref{sigmadef} was chosen to clarify the mathematics, as seen in Enge's ``Deflecting Magnets." Hence, the integrals are equivalent.}, defined as
\begin{equation}
I_2 = \int_{-\sigma_1}^{\sigma_2} \frac{B_y (B_0 - B_y)}{B_0^2} d \sigma.
\end{equation}
This angle $\beta_v$ is the angle through which the pole face in the sharp cutoff case must be rotated to achieve the same vertical effect. 
Although many texts include further approximations to solve for $\beta_v$,\footnote{These texts include but are not limited to SLAC-75, \emph{The Optics of Charged Particle Beams}, and both the TRANSPORT/TURTLE manuals.} no further approximations are necessary. 

Note that Eq.~\eqref{finaldy} is constant in $\sigma$. Therefore, integrating it to find the change in $y$ would only yield second order, trivial terms (from \appr{paraxial approximation} and \appr{thin lens}). Therefore, to first order, $y= y_2 = y_0$. The final effect in the $y$ direction is then
\begin{align}
y &= y_0 \notag \\
y' &= y'_0 - \frac{y_0}{\rho} \left( \tan \beta - \frac{g}{\rho} \frac{1 + \sin^2 \beta}{\cos^3 \beta} I_2 \right).
\end{align}
Within the TRANSPORT code, $R_{43}$ reads
\begin{equation}
R_{43} = -\frac{1}{\rho} \tan \beta_v.
\end{equation}

\subsection{Total Transformation}
Here we compile all effects of the extended fringe field, represented mathematically through matrix-vector algebra. The input vector is equivalent to the output vector from the bending magnet transformation preceding the exit face fringing field. 
\begin{equation}
\left( \begin{array}{c}
x \\ x' \\ y \\ y' 
\end{array} \right) =
\left( \begin{array}{cccc} 
1 & 0 & 0 & 0 \\ 
\frac{1}{\rho} \tan \beta & 1 & 0 & 0 \\ 
0 & 0 & 1 & 0 \\
0 & 0 & -\frac{1}{\rho} \tan \beta_v & 1 \\
 \end{array} \right)
\left( \begin{array}{c}
x_0 \\ x'_0 \\ y_0 \\ y'_0
\end{array} \right) + 
\left( \begin{array}{c}
\frac{g^2}{\rho \cos^2 \beta} I_1 \\ 0 \\ 0 \\ 0
\end{array} \right)
\end{equation}

\section{Evaluation of Approximation Accuracy}

We now consider some real-world magnets and the level of accuracy that the approximations used throughout this document would provide. Two distinct cases are examined: Fermilab's Test Beam dipole magnet and ORKA's first bending magnet (proposed). 

\begin{center}
    \begin{tabular}{ | l | l | l | p{5cm} |}
    \hline
    Value & Test Beam & ORKA Beam & Notes  \\ \hline
    $L$ & 3.0 m & .80 m & Magnet longitudinal length. \\ \hline
    $g$ & 3.8 cm & 8.9 cm & Vertical pole gap - represents twice the maximum vertical displacement. \\ \hline
    $w$ & 13 cm & 30 cm & Magnet width - represents twice the maximum horizontal displacement. \\ \hline
    $p$ & 120 GeV/$c$ & .60 GeV/$c$ & Design momentum of the magnet. \\ \hline
    $B_0$ & 15 kG & 15 kG & Magnitude of the magnetic field deep within the magnet. \\ \hline
    $w/\rho$ & $4.8 \times 10^{-4}$ & .22 & Used in \appr{paraxial approximation}. \\ \hline
    $g/\rho$ & $1.4 \times 10^{-4}$ & .067 & Used in both \appr{paraxial approximation} for twice the maximum vertical displacement and in \appr{gap size}. \\ \hline
    $g/L$ & .013 & .11 & This ratio relates the fringe field length (on the order of $g$) and the total magnet length, used in \appr{thin lens}.\\ \hline
    $L/\rho $ & .011 & .60 & Used in \appr{gap size}. A smaller value yields a  more accurate approximation. \\ \hline
    \end{tabular}
\end{center}

The ratios calculated in the final four rows of the above table are all used in important approximations within this document. If these values are too large (i.e. on the order of $10^{-1}$ or larger), the approximations lose validity because the squares and higher powers of these ratios must be essentially zero. One would note that, upon squaring each of these ratios, the errors from these approximations would be insignificant for a magnet like those in the Test Beam. However, the values yielded by ORKA's dipole are far too large; the squares of these values are still significant.

\section{Conclusion}
The Total Transformation section gives a complete matrix description of a particle's trajectory while passing through the exit fringe field of a dipole magnet. The transfer operations applied here are precisely the same as those found in TURTLE and TRANSPORT. Several other texts including the TRANSPORT manual state a result differing from the result given above (specifically, the $R_{43}$ component). However, the discrepancy lies in the use of an unnecessary approximation to yield a more aesthetically pleasing matrix element.

In conclusion, the first order approximations of the fringe fields in a magnetic dipole are sufficient for high-momentum, large-ring dipoles. Further consideration regarding higher-order effects is necessary when dealing with certain magnets.

\section*{Acknowledgements}
We would like to thank Tom Kobilarcik and Douglas Jensen for their help and advisory throughout the creation of this document. Further, we would not be in any position to write this document without the assistance and encouragement of our advisor Joseph Comfort. Without their aid in revision and experience in the field, this document would not exist.


\begin{thebibliography}{1}

\bibitem{dave} Carey, David C. \emph{The Optics of Charged Particle Beams.} Harwood Academic Publishers, 1992. 

\bibitem{proposal} Fermilab Proposal P1021. \emph{ORKA: Measurement of the $K^+ \rightarrow \pi^+ \nu \overline{\nu}$ Decay at Fermilab.} Fermilab, November 28, 2011.  

\bibitem{enge} Enge, Harald. \emph{Focusing of Charged Particles, Vol. 2.} ``Deflecting Magnets." 

\bibitem{theis} Sagalovsk, Leonid. ``Third-Order Charged Particle Beam Optics." University of Illinois, 1989.


\end{thebibliography}
\end{document}